# Measurements of effective delayed neutron fraction in a fast neutron reactor using the perturbation method[*]

ZHOU Hao-Jun(周浩军)[1, 2; 1)]   YIN Yan-Peng(尹延朋)[2]   FAN Xiao-Qiang(范晓强)[2]

LI Zheng-Hong(李正宏)[2]   PU Yi-Kang(蒲以康)[1]

1 Department of Engineering Physics, Tsinghua University, Beijing, China, 100084

2 Institute of Nuclear Physics and Chemistry, China Academy of Engineering Physics, Mianyang, Sichuan, China, 621900

**Abstract:** A perturbation method is proposed to obtain the effective delayed neutron fraction $\beta_{eff}$ of a cylindrical highly enriched uranium reactor. Based on reactivity measurements with and without a sample at a specified position using the positive periodic technique, the reactor reactivity perturbation $\Delta\rho$ of the sample in $\beta_{eff}$ units is measured. Simulations of the perturbation experiments are performed using the MCNP program. The PERT card is used to provide the difference $dk$ of effective neutron multiplication factors with and without the sample inside the reactor. Based on the relationship between the effective multiplication factor and the reactivity, the equation $\beta_{eff}=dk/\Delta\rho$ is derived. In this paper, the reactivity perturbations of 13 metal samples at the designable position of the reactor are measured and calculated. The average $\beta_{eff}$ value of the reactor is given as 0.00645, and the standard uncertainty is 3.0%. Additionally, the perturbation experiments for $\beta_{eff}$ can be used to evaluate the reliabilities of the delayed neutron's parameters. This work showed that the delayed neutron data of $^{235}$U and $^{238}$U from G.R. Keepin's publication are more reliable than those from ENDF-B6.0, ENDF-B7.0, JENDL3.3 and CENDL2.2.

**Key words:** Effective delayed neutron fraction, reactivity perturbation, fast neutron reactor

**PACS:** 28.20.-v, 28.50.Ft, 29.90.+r

## 1 Introduction

The effective delayed neutron fraction $\beta_{eff}$ is an important parameter in reactor physics. It is often used as a unit of experimental reactivity in dollar ($, 1$=100¢), and the contribution of delayed neutrons is of primary importance for control rod worth calculation, reactivity feedback effect studies and reactivity accident analysis. A more accurate $\beta_{eff}$ can also be used to validate the delayed neutron data of $^{235}$U, $^{238}$U and $^{239}$Pu.

The Cylindrical Highly Enriched Uranium Reactor (CHEUR), operated by the Institute of Nuclear Physics and Chemistry, China Academy of Engineering, is a fast neutron reactor specified for neutron physics studies, fast reactor benchmark experiments and neutron application techniques. In order to improve the $\beta_{eff}$ measurement of CHEUR, a modified method is proposed based on the $^{252}$Cf source method [1]. The $^{252}$Cf source method has been applied so far to the $\beta_{eff}$ benchmark measurements [2, 3]. The reactor must be in subcritical state, and $\beta_{eff}$ is determined by the ratio of the experimental reactivity in dollar and the calculated result ($1-1/k_{eff}$). Generally, the subcritical reactivity measurements are less accurate than the super delayed critical reactivity experiments. The

[*] Supported by Foundation of Key Laboratory of Neutron Physics, China Academy of Engineering Physics (Contract No: 2012AA01, 2014AA01).

[*] Supported by The National Natural Science Foundation (Contract No: 11375158, 91326104).

1) E-mail: Vampirl@163.com

calculated effective multiplication factor $k_{eff}$ needs revision due to the difference between the simulation model and the real reactor. For the modified method, the subcritical reactivity measurements are replaced with the reactivity perturbation experiments with a super delayed critical state. Furthermore, the simulation results of the perturbation experiments do not need any revisions. Because perturbation theory is used in the experimental simulation, we name this the perturbation method.

Section 2 gives a description of the perturbation method. Section 3 and 4 give a brief introduction of CHEUR and the location of the perturbed samples. Section 5, 6 and 7 give a detailed description of the reactivity perturbation experiments and simulations, and the result and uncertainties of $\beta_{eff}$ are discussed. In Section 8, the perturbation experiments for $\beta_{eff}$ are used to evaluate the reliability of the delayed neutron data.

## 2 The perturbation method

Assuming that the reactor is in a super delayed critical state, the power growth period $T_1$ of the reactor can be obtained by the positive period method. The reactivity $\rho_1$ in dollar is well described as follows.

$$\rho_1 = \frac{1}{\beta_{eff}} \frac{\Lambda}{T_1} + \sum_{q} \sum_{i=1}^{I} \frac{a_i^q}{1 + \lambda_i^q T_1} \quad \ldots\ldots(1)$$

where $\Lambda$ is the neutron generation time of the reactor and $a_i^q$, $\lambda_i^q$ are the fractions and decay constants respectively of the precursor of the $i^{th}$ group of delayed neutrons for a certain nuclide $q$.

When a small reactivity perturbation $\Delta\rho$ is introduced into the reactor, the power growth period $T_2$ of the reactor is obtained in the same way. The reactivity $\rho_2$ is given by the following equation.

$$\rho_2 = \frac{1}{\beta_{eff}} \frac{\Lambda}{T_2} + \sum_{q} \sum_{i=1}^{I} \frac{a_i^q}{1 + \lambda_i^q T_2} \quad \ldots\ldots(2)$$

The reactivity perturbation $\Delta\rho$ is received by subtracting $\rho_1$ from $\rho_2$. Because the neutron generation time $\Lambda$ is 9 orders of magnitude smaller than the power growth period of the reactor, terms including $\Lambda$ are neglected, as shown in Eq. 3.

$$\Delta\rho = \sum_{q} \sum_{i=1}^{I} \left( \frac{a_i^q}{1 + \lambda_i^q T_2} - \frac{a_i^q}{1 + \lambda_i^q T_1} \right) \quad \ldots\ldots(3)$$

A simulation has been implemented to describe the process mentioned above. The difference $dk$ between the effective multiplication factors with and without the perturbation can be calculated using the deterministic method or the Monte Carlo method. Based on the relationship of the reactivity and the effective multiplication factor, the $dk$ value can be directly used to express the reactivity perturbation when the reactor is close to a delayed critical state. Since $dk$ and $\Delta\rho$ denote the same physical parameter and $\beta_{eff}=dk/\Delta\rho$, $\beta_{eff}$ is rewritten in the following equation.

$$\beta_{eff} = \frac{dk}{\Delta\rho} = \frac{dk}{\sum_{q} \sum_{i=1}^{I} \left( \frac{a_i^q}{1 + \lambda_i^q T_2} - \frac{a_i^q}{1 + \lambda_i^q T_1} \right)} \quad \ldots\ldots(4)$$

## 3 Configuration of the reactor

CHEUR is a cylindrical fast neutron reactor. The structure of the core is composed of a control rod, a lower active zone, a middle steel disc and an upper active zone. The upper active zone is a highly enriched uranium (HEU) cylinder reflected by natural uranium. On the top of the HEU cylinder, an 11 mm-thick disc is disassembled into four reactivity adjustment components (i.e. A, B, C and D, shown in Fig.1). Another four components have been manufactured in the same dimensions but with stainless steel metal. The reactor reactivity can be controlled by removing one adjustment component or changing its material. Similar to the upper counterpart, the lower active zone is also an HEU cylinder reflected by natural uranium. In the center of the lower active zone, only one HEU metal control rod is used during the operation of the reactor.

## 4  Experimental setup

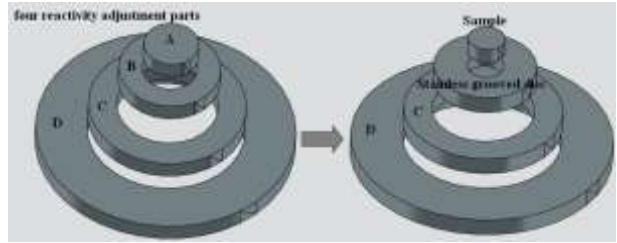

**Fig. 1 Location of the perturbed samples in the reactor**

To simplify the introduction of the perturbed samples, the experiments are fitted right at the top of the reactor. As shown in Fig. 1, the reactivity adjustment components A and B were removed and replaced by the stainless steel disc. The outer diameter of the disc is 52.00 mm and the height is 11.00 mm. In the center of the disc, there is a groove 20.00 mm in diameter and 9.00mm in depth, where the experimental samples can be placed.

**Table 1 Mass and dimensions of the samples**

| Samples | Diameter/mm | Height/mm | Mass/g | Volume/mm$^3$ | Density/(g/cm$^3$) |
|---|---|---|---|---|---|
| Au | 19.96 | 9.00 | 54.2825 | 2816.1349 | 19.2755 |
| Fe | 19.98 | 9.00 | 22.1582 | 2821.7813 | 7.8526 |
| Ni | 19.96 | 9.08 | 25.3080 | 2841.1672 | 8.9076 |
| Ti | 19.96 | 9.04 | 12.7276 | 2828.6511 | 4.4995 |
| Cu | 19.98 | 9.04 | 25.2217 | 2834.3226 | 8.8987 |
| V | 19.96 | 9.06 | 17.2074 | 2834.9091 | 6.0698 |
| Ag | 19.98 | 8.96 | 29.4497 | 2809.2401 | 10.4832 |
| Bi | 19.98 | 9.00 | 27.6326 | 2821.7813 | 9.7926 |
| Zr | 19.98 | 8.98 | 18.5068 | 2815.5107 | 6.5732 |
| Pb | 19.94 | 8.96 | 31.6016 | 2798.0031 | 11.2943 |
| Cr | 19.98 | 9.00 | 20.1186 | 2821.7813 | 7.1298 |
| Al | 19.96 | 8.96 | 7.5375 | 2803.6188 | 2.6884 |
| Cd | 19.98 | 9.00 | 24.2944 | 2821.7813 | 8.6096 |

Thirteen neutron scattering samples, including Au, Fe, Ni, Ti, Cu, V, Ag, Bi, Zr, Pb, Cr, Al and Cd, were prepared for the reactivity perturbation experiments. These are naturally enriched materials and the impurity of each sample was less than 0.1%. Raw materials were provided by the Central Research Institute for Nonferrous Metals and were manufactured into Φ19.98 mm×9.00 mm cylindrical samples.

Before the perturbation experiments, a vernier caliper (0.02 mm in precision) and the high-precision electronic scale (120 g in measuring range and 0.1 mg in precision) were used to get the accurate dimensions and masses of all prepared samples. The measurement results are given in Table 1. Additionally, the volume and density of each were obtained for the following experimental simulations.

## 5  Reactivity measurements

The operating conditions of CHEUR during the experiments were as follows. The reactivity adjustment component D was set to be HEU; component C was stainless steel; components A and B were replaced by the stainless steel grooved disc. When the lower active zone is in tight contacts with the middle steel disc and the control rod reaches the inner limit by inserting the rod body 95.00 mm into the lower active zone, the reactor goes into a super delayed critical state. The reactivity is about 0.12$.

Initially the sample grove was empty. The lower active zone was in direct contacts with the middle steel disc, and the control rod reached the inner limit. After CHEUR went into the super delayed critical state, the positive periodic meter was then used for the power growth period $T_1$. The reactor was shut down and a sample placed in the groove. Similar to the previous approach, once the lower active zone made contact with the middle steel disc, the control rod reached the inner limit a second time. As long as we had measured the power growth period $T_2$, the reactor was shut down temporally and another sample was used to replace the original sample in the sample groove. This procedure is repeated while other operating conditions were kept the same until all the samples were measured. In operations, the power range of the reactor was within 0.5~5 W during each power growth period. If the power level is too low, the current signal from the detectors will not be intense enough, leading to serious compromises in resolution. If the power level is too large, the temperature of the reactor increases significantly due to fission heat generation from nuclear burnup.

**Table 2 Reactivity results measured in November, 2013**

| Samples | Power growth periods/s | Reactivity/¢ | Temperatures/℃ |
| --- | --- | --- | --- |
| Empty | 51.69 | 12.29 | 15.5 |
| Fe | 35.07 | 16.14 | 15.5 |
| Ni | 31.53 | 17.32 | 15.5 |
| Ti | 39.16 | 14.96 | 15.5 |
| Cu | 31.92 | 17.18 | 15.5 |

All the measurements were performed within two time periods. The first operating period was in November, 2013. The reactor reactivity with no sample and with Fe, Ni, Ti, Cu as the sample are recorded in Table 2. The second operating period was in March, 2014. The reactor reactivity with and without the sample in the sample groove were measured and are listed in Table 3. All the reactivity results were derived based on the delayed neutron data presented by G.R. Keepin [4].

**Table 3 Reactivity results measured in March, 2014**

| Samples | Power growth periods/s | Reactivity/¢ | Temperature/℃ |
|---------|------------------------|--------------|---------------|
| Empty   | 48.42                  | 12.87        |               |
| Au      | 29.52                  | 18.90        |               |
| V       | 31.58                  | 17.30        |               |
| Ag      | 32.04                  | 17.14        |               |
| Bi      | 36.65                  | 15.66        | 13.5          |
| Zr      | 32.85                  | 16.85        |               |
| Pb      | 35.66                  | 15.95        |               |
| Cr      | 32.56                  | 16.84        |               |
| Al      | 35.94                  | 15.87        |               |
| Cd      | 34.38                  | 16.36        |               |

# 6 $\beta_{eff}$ results

For a specified sample, the difference between the reactivity with and without the sample in the sample groove can be calculated based on the data in Tables 2 and 3. The reactivity perturbation results of various samples are listed in Table 4.

The simulation of the perturbation experiments on CHEUR was performed with the MCNP program. The PERT card was used to calculate the difference of the effective multiplication factors, *dk*, of CHEUR in scenarios with and without a sample in the groove. According to *dk* and *Δρ* given in Fig. 2 and the least squares fitting method, the effective delayed neutron fraction of CHEUR is calculated as 0.00645.

**Table 4 Reactivity perturbation of various samples**

| Samples | Fe | Ni | Ti | Cu | Au | V | Ag | Bi | Zr | Pb | Cr | Al | Cd |
|---------|------|------|------|------|------|------|------|------|------|------|------|------|------|
| Reactivity Perturbation /¢ | 3.87 | 5.06 | 2.74 | 4.97 | 5.22 | 4.42 | 4.27 | 2.79 | 3.98 | 3.08 | 4.00 | 3.04 | 3.54 |

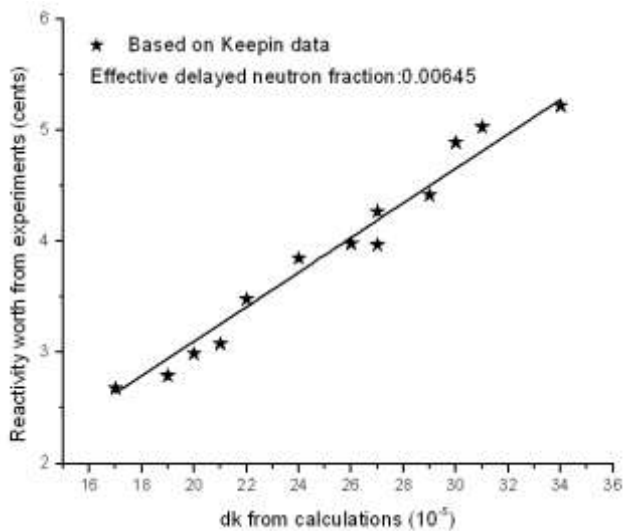

**Fig. 2 The effective delayed neutron fraction based on G.R. Keepin's data**

## 7 Uncertainty analysis

Represent the relative uncertainties of $\Delta\rho$ and $dk$ as $u_\rho$ and $u_k$, respectively. The relative uncertainty $u_\beta$ of the $\beta_{eff}$ value can be described by Equation (5). Due to the independence between the experiments and theoretical calculations, there is no interrelationship between $u_\rho$ and $u_k$ in the formula.

$$u_\beta^2 = u_\rho^2 + u_k^2 \quad\quad\quad\quad\quad\quad\quad\quad\quad\quad\quad\quad\quad\quad\quad\quad\quad\quad\quad\quad\quad\quad\quad\quad\quad\quad (5)$$

One of the major sources of $u_\rho$ is the non-reproducibility of the reactor operations. In the experiments, the reactivity perturbation of a sample is obtained by measuring the difference between the reactivity with and without a sample in the sample groove. The process of placing a sample in the reactor core requires shutting down the reactor. Due to the nature of randomness, the contact level between the lower active zone and the steel disc and the positions of the control rod can never be perfectly identical between two operations.

Under the same operating conditions, as mentioned above in the perturbation experiments, the lower active zone makes contact with the middle steel disc and the control rod reaches the inner limit. Once the power growth period of the reactor is obtained by the periodic meter, the reactor is shut down immediately. When the reactor power decreases to less than 0.01 W, the procedures are repeated until ten reactivity measurements are completed. The non-reproducibility of the reactor operations can be expressed by twice Bessel standard deviation of the average reactivity results. The non-reproducibility result of this experiment is 0.036 ¢.

The worst scenario during the perturbation experiments is that the working conditions for reactivity measurements were unfortunately at the both extreme end of the statistical distribution of the reactor's non-reproducibility. In order to correct the undesirable conditions, the presumed reactor's non-reproducibility should be doubled to 0.072 ¢.

The uncertainty of the delayed neutron parameter is another major source of $u_\rho$. As mentioned above, the reactivity perturbation results are calculated based on the Keepin's delayed neutron parameters. The uncertainties for these provided in reference [4] are introduced in the $\Delta\rho$ measurements. The influence of the uncertainties of Keepin's data to $u_\rho$ is 0.04 ¢.

The effects of the reactor non-reproducibility are independent of those of the delayed neutron parameters.

The $\Delta\rho$ measurement uncertainty can be evaluated based on the square root of the square sum of the two values, i.e. $\pm 0.08\ ¢$. For the measurement results of the reactivity perturbation given in Table 4, the relative uncertainty $u_\rho$ is within 1.6%~2.6%.

$dk$ values were obtained with the MCNP program by running the PERT card, and the sources of the uncertainties $u_k$ consist of the calculation methods, the difference between the established model and the reactor operating in reality, and the nuclear database used in the calculation. Making calculations using slightly modified criticality of the reactor model and different nuclear databases, the results show not much change. The main contribution of $u_k$ can be attributed to the calculation method itself. In this paper, the data of uncertainty analysis for the PERT card presented in reference [5] has been adopted. The PERT perturbation estimator typically provides sufficient accuracy for response or tally changes that are less than 5%.

In summary, the $\beta_{eff}$ values of CHEUR were measured by the perturbation method, and the relative uncertainty of a single measurement is 5.6% based on Equation (5). The main uncertainty source of the result rises from the PERT calculation method. The final $\beta_{eff}$ result is given by the least squares method on the reactivity perturbation measurements of 13 samples. Bessel standard deviation of the least squares fitting is 4.2%. Assuming that the probability distribution of the single measurement results in the uncertainty range is normal distribution, the confidence factor is 2. According to the method suggested in reference [6], the relative uncertainty $u_c$ of $\beta_{eff}$ is written in Equation (6).

$$u_c = \sqrt{\frac{(5.6\%)^2}{4} + \frac{(4.2\%)^2}{13}} = 3.0\% \quad \cdots\cdots(6)$$

## 8 Delayed neutron parameters validation

Nowadays, there exist multiple versions of the delayed neutron group parameters. Five fast fission delayed neutron parameters [7] of $^{235}$U presented in Keepin's work, ENDF-B6.0, ENDF-B7.0, JENDL3.3 and CENDL2.2 are given in Table 5. The difference among them is obvious. The perturbation method gives $\beta_{eff}$, and seems to provide a feasible way for validating the reliability of delayed neutron parameters.

The six-group delayed neutron parameters of $^{235}$U and $^{238}$U from ENDF-B6.0, ENDF-B7.0, JENDL3.3 and CENDL2.2 database are used to replace those from Keepin's work. The $\beta_{eff}$ values are given in Fig. 3 (0.00710), Fig. 4(0.00730), Fig. 5(0.00621) and Fig. 6 (0.00717) respectively.

For the purpose of validating the reliability of delayed neutron parameters, it is necessary to verify the $\beta_{eff}$ value of CHEUR using another method.

As reported in reference [4], the $\beta_{eff}$ values of $^{235}$U and $^{238}$U are 0.00640 and 0.01480, respectively. In reference [8], the $\beta_{eff}$ values of Godiva (HEU-MET-FAST-001 from ICSBEP) and Flattop-25 (HEU-MET-FAST-028 from ICSBEP) are 0.00659 and 0.00665, respectively. Godiva is a spherical HEU nuclear facility, and Flattop-25 is a spherical HEU facility with a normal uranium reflector outside. In our scenario, CHEUR is also a HEU facility with a natural uranium reflector. However, the mass of the reflector used in CHEUR is less than tenth of the reflector in Flattop-25. In accordance with the approximation principle, the $\beta_{eff}$ value of CHEUR is presumably higher than that of Godiva but lower than that of Flattop-25. Thus, considering the $\beta_{eff}$ measurement uncertainties, we can reasonably predict that the $\beta_{eff}$ value of CHEUR is most possibly

within 0.00640~0.00680.

**Table 5 Six-group delayed neutron parameters of $^{235}$U fast fission**

| $i$ | $\lambda_i/\text{s}^{-1}$ | | | | | $\alpha_i$ | | | | |
|---|---|---|---|---|---|---|---|---|---|---|
| | Keepin | ENDFB6.0 | ENDFB7.0 | JENDL3.3 | CENDL2.2 | Keepin | ENDFB6.0 | ENDFB7.0 | JENDL3.3 | CENDL2.2 |
| 1 | 0.0127 | 0.0133 | 0.0125 | 0.0124 | 0.0133 | 0.038 | 0.0350 | 0.0320 | 0.035 | 0.0350 |
| 2 | 0.0317 | 0.0327 | 0.0318 | 0.0305 | 0.0347 | 0.213 | 0.1807 | 0.1664 | 0.217 | 0.1807 |
| 3 | 0.115 | 0.121 | 0.109 | 0.111 | 0.121 | 0.188 | 0.1725 | 0.1613 | 0.212 | 0.1725 |
| 4 | 0.311 | 0.303 | 0.317 | 0.301 | 0.305 | 0.407 | 0.3868 | 0.4596 | 0.385 | 0.3868 |
| 5 | 1.40 | 0.849 | 1.354 | 1.136 | 0.849 | 0.128 | 0.1586 | 0.1335 | 0.126 | 0.1586 |
| 6 | 3.87 | 2.853 | 8.636 | 3.014 | 2.855 | 0.026 | 0.0664 | 0.0472 | 0.025 | 0.0664 |

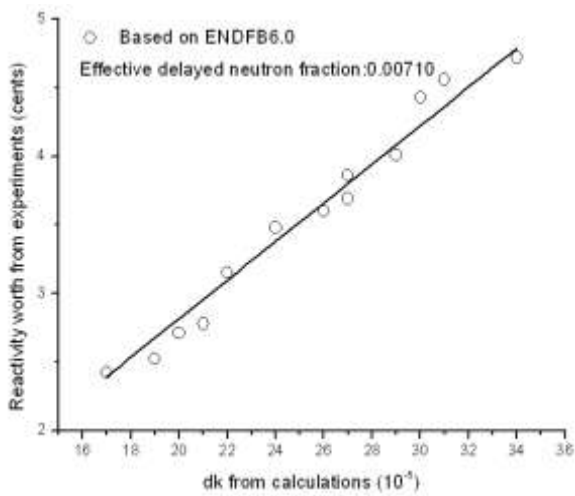

**Fig. 3** $\beta_{eff}$ based on the delayed neutron parameters from ENDF-B6.0

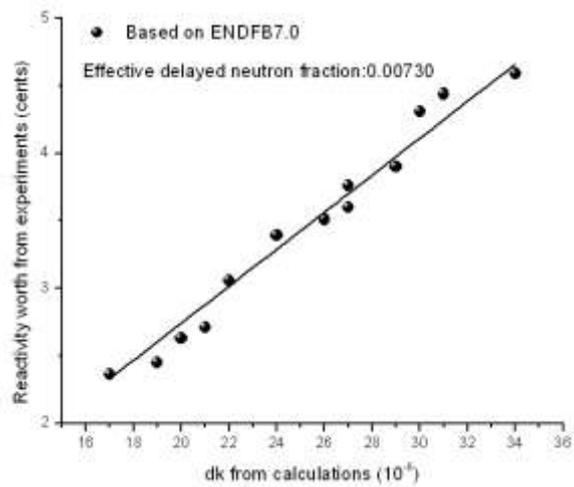

**Fig. 4** $\beta_{eff}$ based on the delayed neutron parameters from ENDF-B7.0

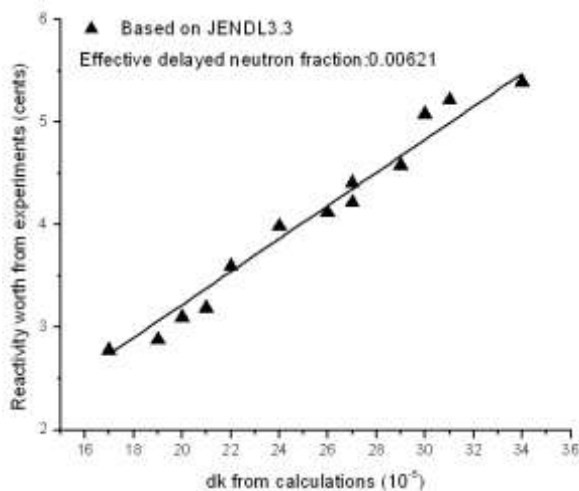

**Fig. 5** $\beta_{eff}$ based on the delayed neutron parameters from JENDL3.3

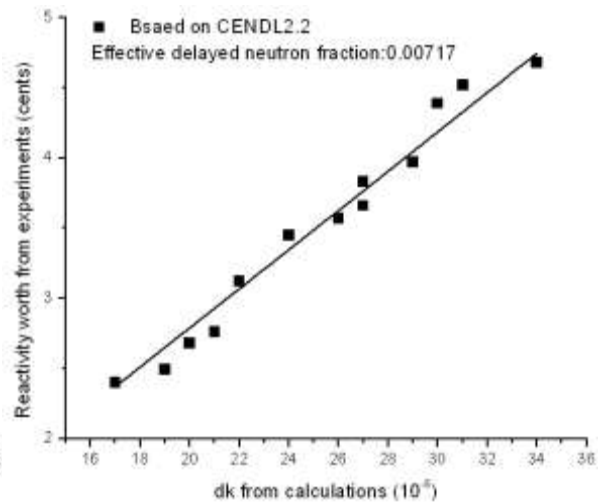

**Fig. 6** $\beta_{eff}$ based on the delayed neutron parameters from CENDL2.2

The $β_{eff}$ of CHEUR was calculated by the $k$-ratio method [9] using the MCNP program with the ENDF-B6.6 database. The effective neutron multiplication factors ($k_{eff}$ and $k_p$) with and without considering the delayed neutrons were calculated. Based on the equation $β_{eff} = 1 - k_p/k_{eff}$, the effective delayed neutron fraction $β_{eff}$ was obtained as 0.00667.

It is found that only $β_{eff}$ calculated based on G.R. Keepin's delayed neutron parameters is within the specific range predicted by the approximation principle and close to the theoretical calculation value given by the $k$-ratio method. This indicates that the six-group delayed neutron parameters of $^{235}$U and $^{238}$U provided by G.R. Keepin are more reliable than the other data listed in Table 5.

# 9 Conclusions

The perturbation method has been proposed to measure the effective delayed neutron fraction $β_{eff}$ in CHEUR. According to reactivity measurements with and without a sample at the specified position by the positive periodic technique, the reactivity perturbation $Δρ$ of the sample in $β_{eff}$ units was achieved. Further, simulation of the perturbation experiment was performed using the MCNP program. The PERT card was used to provide the difference $dk$ of effective neutron multiplication factors with and without a sample in the reactor. Comparing the experimental result with the calculated result, the equation $β_{eff} = dk/Δρ$ was found.

In this work, the reactivity perturbations of 13 different metal samples at the specified position of the reactor were measured and calculated, and the average $β_{eff}$ value of the reactor was found to be 0.00645. The relative uncertainty of the $β_{eff}$ result is 3.0%. The main sources of the uncertainty consist of the non-reproducibility of the reactor operations, the uncertainties of the delayed neutron parameters, and the perturbation calculating methods.

Since the $β_{eff}$ measurement with perturbation method is highly sensitive to the delayed neutron parameters, it can be used to evaluate the reliability of the delayed neutron parameters. This paper shows that the delayed neutron parameters of $^{235}$U and $^{238}$U coming from G.R. Keepin's work are more reliable than those coming from ENDF-B6.0, ENDF-B7.0, JENDL3.3, and CENDL2.2.